\documentclass[runningheads,a4paper]{llncs}

\usepackage{amssymb}
\usepackage{graphicx}

\usepackage{url}
\urldef{\mailsa}\path|{luis.sanabria, jaume.barcelo, albert.domingo, boris.bellalta}@upf.edu|

\newcommand{\keywords}[1]{\par\addvspace\baselineskip
\noindent\keywordname\enspace\ignorespaces#1}

\begin{document}
\mainmatter
\title{Spectrum Sensing with USRP-E110}
\author{Luis Sanabria-Russo\and Jaume Barcelo\and Albert Domingo\and Boris Bellalta}
\institute{Universitat Pompeu Fabra\\
	   Carrer de T\`anger $122-150$, 08018 Barcelona, Spain.\\
  \mailsa\\}

\maketitle

\begin{abstract}

Spectrum sensing is one of the key topics towards the implementation of future wireless services like SuperWiFi. This new wireless proposal aims at using the freed spectrum resulting from the analog-to-digital transition of TV channels for wireless data transmission (UHF TV White Spaces). The benefits range from better building penetration to longer distances when compared to the set of IEEE 802.11 standards. Nevertheless, the effective use of the available spectrum is subject to strict regulation that prohibits unlicensed users to interfere with incumbents (like wireless microphones). Cognitive Radios (CR) and dynamic spectrum allocation are suggested to cope with this problem. These techniques consist on frequency sweeps of the TV-UHF band to detect White Spaces that could be used for SuperWiFi transmissions. In this paper we develop and implement algorithms from GNURadio in the Ettus USRP-E110 to build a standalone White Spaces detector that can be consulted from a centralized location via IP networks.

\end{abstract}

\section{Introduction}
  TV White Spaces refer to free spectrum available in the TV band. These range from VHF to UHF bands, although in this paper only those in UHF will be considered (from $471.25$ to $863.25$ MHz). The switchover from analog to digital television increased the number of TV White Spaces, which are going to be used for upcoming wireless standards, like IEEE 802.22 and SuperWiFi.

The benefits of attempting data transmission over these frequencies include longer ranges and better building penetration compared with the unlicensed $2.4$\  GHz and $5$\  GHz bands used in the set of IEEE 802.11 protocols. Along with these benefits, a variety of technical and regulatory challenges surface, involving spectrum sensing, channel availability and incumbent avoidance.

The technical difficulties involving spectrum sensing vary with each technique~\cite{shellhammer2009technical}. Some of the more popular are energy detection~\cite{energyDetection}\cite{techChallengesUSRP}, cyclostationary detection~\cite{cycloDetection} and locking detection~\cite{lockDetection}.

In this work, an energy detection-based approach is implemented due to its low complexity. It is built using a standalone Universal Software Radio Peripheral (USRP) Ettus USRP-E110~\cite{USRPE110} and the open source Software Defined Radio (SDR) project GNURadio~\cite{GNURadio}. USRP-E110 is a popular SDR platform that acts as a radio front-end for embedded applications. Furthermore, it is equipped with the USRP Hardware Driver (UHD)~\cite{UHD} that allows the writing of GNURadio programming code at a separate host computer and then to transfer it to the USRP-E110 via a standard Secure Shell (SSH) tunnel (see the connections layout in Fig.~\ref{fig:connections}).

Results show a clear image of the studied spectrum that provides the opportunity to recognize unused TV channels. 

Apart from signal processing, the combination USRP-GNURadio has a steep learning curve that requires familiarity with Python and C/C++ programming languages. 

In Sect.~\ref{Section2} the connection layout between a PC and the USRP-E110 is described. A proposal for identification and selection of White Spaces in a standalone Ettus USRP-E110 is presented in Sect.~\ref{Section3}. Section~\ref{Section4} summarizes the results of this paper and highlights the future work pending in this area.
  
\section{Spectrum Sensing} \label{Section2}
  TV channels are 6 and 8 MHz wide in USA and Europe respectively. TV White Spaces, which are of the same bandwidth, are shared with other transmitters like microphones and radio astronomy. The main objective of an effective spectrum sensing is to avoid the channels where these incumbents are transmitting.

In spite of the technical challenges, there are several techniques available to aid the task of finding an empty channel. The proposed approach gathers the Power Spectral Density (PSD) of a set of samples taken every SampleStep (SS) Hertz in a determined frequency band. After the gathering, post-processing carries the task of identifying which power levels could be considered as noise based on the statistics of the sampled spectrum.

\begin{figure}[htbp]
  \centering
  \includegraphics[width = 9cm]{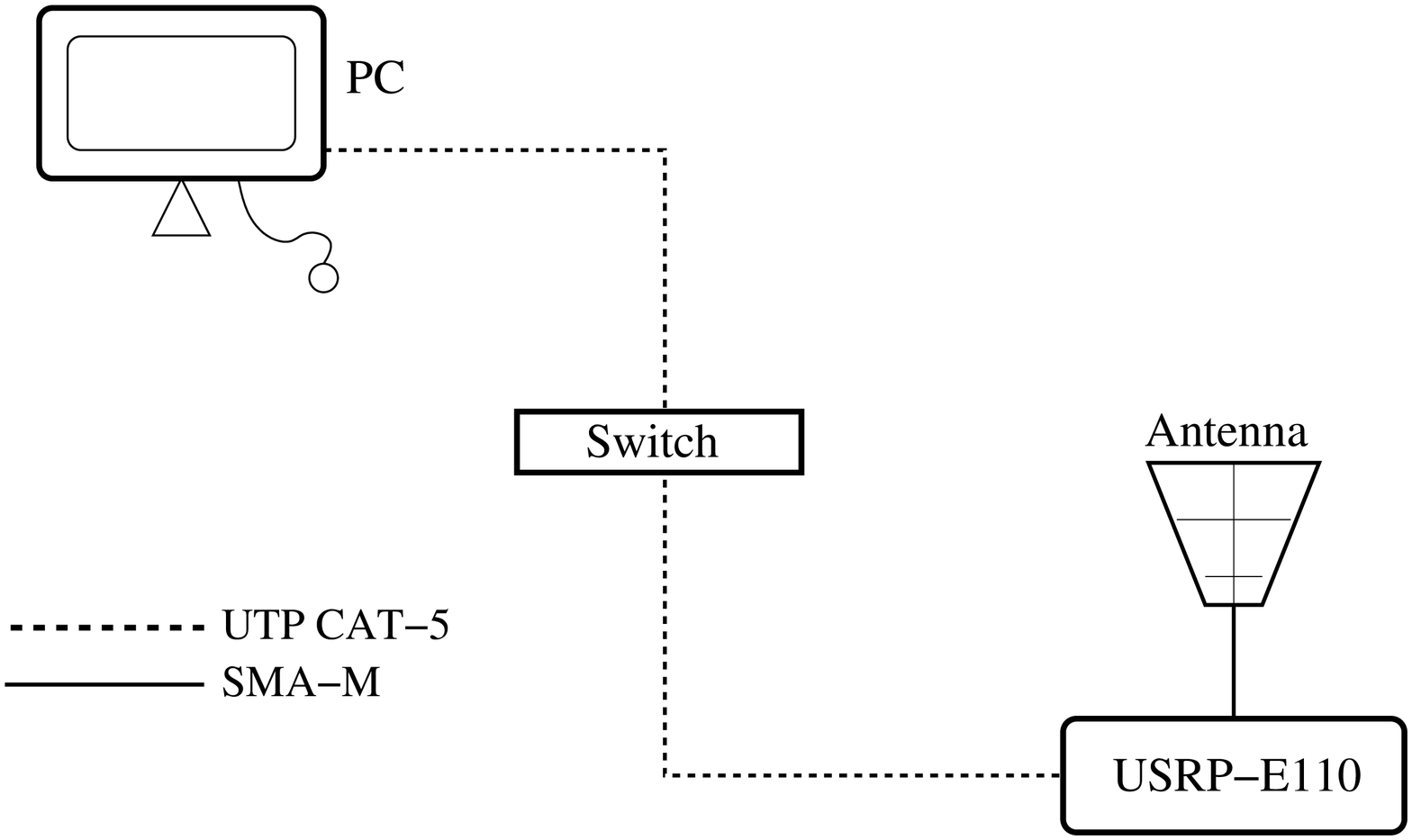}
  \caption{PC-USRP connections layout}
  \label{fig:connections}
\end{figure}

The identification of White Spaces is performed in a standalone USRP-E110 using GNURadio~\cite{GNURadio} and a modified example code for spectrum sensing ~\cite{uhdSpecSense}\cite{code} to account for different RF daughterboards and antennas. Figure~\ref{fig:connections} presents the connection layout between the USRP-E110 and a PC. This connection was used to transmit the executable code to the USRP and then to retrieve the White Spaces estimation file.

Both the PC and USRP-E110 belong to the same subnet and the communication was established inside a standard SSH tunnel.

  
\section{Using the USRP-E110 as a standalone White Spaces detector} \label{Section3}
  The USRP is equipped with an Ettus WBX daughterboard~\cite{WBX} that allows the reception of all TV channels in the UHF band. The setup is also composed by a log-periodic directional antenna \cite{logPeriodic} which is connected to the USRP via a SMA-M to SMA-M cable.

\subsection{Spectrum Sense in USRP via UHD}
UHD~\cite{UHD} allows the configuration of the USRP via SSH or USB through its console port. It is installed by default in current versions of the different USRPs models available and provides support for operations like spectrum sensing controlled from a remote location.

In order to gather the signal power in the UHF TV band, the example spectrum sensing code~\cite{uhdSpecSense} is modified to comply with the proposed task \cite{code}. Developing on what was mentioned in Sect.~\ref{Section2}, the algorithm measures the signal power at a center frequency $f_{o}$ for a fixed \emph{dwelling time} $d_{w}=0.001$~s, moving from $f_{min}$ towards $f_{max}$ at $SS$ frequency intervals. We denote the signal power at interval $i$ as $P_i$ and $T=(f_{max}-f_{min}) / SS$ samples are taken in total. Since $SS$ is typically smaller than the width of a TV channel, several samples are taken for each TV channel.

The procedure described above is conventionally performed using a spectrum analyzer. The resolution bandwidth in the spectrum analyzer is equivalent to our $SS$. In our prototype, there is not a concept equivalent to the spectrum analyzer's video bandwidth, as we do not apply any further filtering on the $P_{i}$ values.



\subsection{Identifying TV White Spaces}

After all the signal power readings (or collection) are contained in a file, a threshold ($\gamma$ in~(\ref{eq:threshold})) is defined.

\begin{equation} \label{eq:threshold}
 \gamma = avg(\min{P_{i}},\max{P_{i}})
\end{equation}




If a received power in the collection ($P_i$, $i=1,2,...,T$) is greater than $\gamma$, then the whole channel at which the sample belongs to is considered \emph{occupied}. On the contrary case, it is considered \emph{free} if $P_{i} \le \gamma$. This measure is used in order to avoid channels where narrow-band transmissions might be present (like wireless microphones).

\begin{figure}[htbp]
  \centering
  \includegraphics[width = 8.5cm, angle = -90]{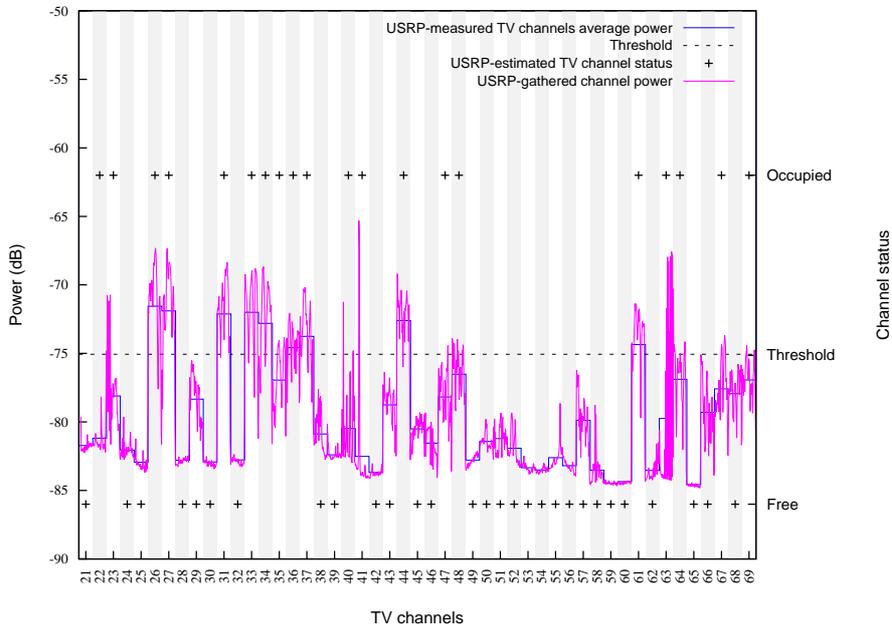}
  \caption{USRP-estimated TV White Spaces}
  \label{fig:tvChannels}
\end{figure}

A graphical representation is found in Fig.~\ref{fig:tvChannels}, where $f_{min}=471.25\  \rm{MHz}$, $f_{max}=863.25\  \rm{MHz}$ and $SS=250\  \rm{kHz}$, resulting in thirty two power samples per TV channel. 





A study performed by Domingo \emph{et al.}~\cite{whiteSpacesCatalonia}, documented a spectrum sweep with a spectrum analyzer aimed at finding TV White Spaces at the same location as the testings in this work. The presented approach matched nearly 70\% of their observations, revealing 29 TV White Spaces against the 35 observed with the spectrum analyzer. This lower number is possibly due to narrow-band transmissions detected by the USRP.
\\

\section{Conclusions and future work} \label{Section4}
  The proposed identification of TV White Spaces with USRP-E110 enables the execution of a spectrum sensing algorithm via SSH, allowing the USRP to be located at convenient locations.

It is possible to build a Radio Environment Maps (REM)~\cite{REM} from samples gathered by geographically distributed USRPs controlled from a centralized location, increasing the efficiency and boosting the implementation of cognitive networks.

In order optimize the spectrum sensing algorithm, better signal processing techniques are expected to be implemented in the near future~\cite{shellhammer2009technical}. Also it is planned to add a second step in the analysis that involves time domain samples gathering in order to perform stochastic analysis of the signal. All of this in the attempt to differentiate noise from TV broadcast signals.

  
 \section*{Acknowledgements} \label{ack}
  This work has been partially funded by the European Commission (grant CIP-ICT PSP-2011-5) and the Spanish Government (TEC2008-0655, Plan Nacional I+D). The views expressed in this work are solely those of the authors and do not represent the views of the European Commission nor of the Spanish Government.

\bibliographystyle{splncs03}

\end{document}